\documentclass[trackchanges,twocolumn,times,usenatbib]{aastex631}

\usepackage{melissa_style}

\newcommand{\of}[1]{{\color{blue}{#1}}}
\definecolor{rust}{rgb}{0.7,0.1,0.1}
\newcommand\N[1]{{\color{rust} \bf #1}} 
\usepackage{etoolbox}
\pretocmd{\abstractname}{\newpage}{}{}

\graphicspath{{./}{Figures/}}

\revised{\today}

\submitjournal{ApJ}
\shorttitle{\jwst\ data of Type~IIn \ip}
\shortauthors{Shahbandeh, et al.}

\begin{document}
\title{\jwst/MIRI Observations of Newly Formed Dust in the Cold, Dense Shell of the Type IIn SN~2005ip}

\author[0000-0002-9301-5302]{Melissa Shahbandeh}
\altaffiliation{STScI Fellow}
\affiliation{Space Telescope Science Institute, 3700 San Martin Drive, Baltimore, MD 21218, USA}

\author[0000-0003-2238-1572]{Ori D.Fox}
\affiliation{Space Telescope Science Institute, 3700 San Martin Drive, Baltimore, MD 21218, USA}

\author[0000-0001-7380-3144]{Tea Temim}
\affiliation{Department of Astrophysical Sciences, Princeton University, Princeton, NJ 08544, USA}

\author{Eli Dwek}
\affiliation{Emeritus, Observational Cosmology Lab, NASA Goddard Space Flight Center, Mail Code 665, Greenbelt, MD 20771, USA}
\affiliation{Research Fellow, Center for Astrophysics — Harvard \& Smithsonian, 60 Garden Street, Cambridge, MA 02138, USA}

\author[0000-0002-9820-679X]{Arkaprabha Sarangi}
\affiliation{DARK, Niels Bohr Institute, University of Copenhagen, Jagtvej 155A, 2200 Copenhagen, Denmark}

\author[0000-0001-5510-2424]{Nathan Smith}
\affiliation{Steward Observatory, University of Arizona, 933 N. Cherry St, Tucson, AZ 85721, USA}

\author[0000-0003-0599-8407]{Luc Dessart}
\affiliation{Institut d'Astrophysique de Paris, CNRS-Sorbonne Universit\'e, 98 bis boulevard Arago, F-75014 Paris, France}

\author{Bryony Nickson}
\affiliation{Space Telescope Science Institute, 3700 San Martin Drive, Baltimore, MD 21218, USA}

\author[0000-0003-0209-674X]{Michael Engesser}
\affiliation{Space Telescope Science Institute, 3700 San Martin Drive, Baltimore, MD 21218, USA}

\author[0000-0003-3460-0103]{Alexei V. Filippenko}
\affiliation{Department of Astronomy, University of California, Berkeley, CA 94720-3411, USA}

\author[0000-0001-5955-2502]{Thomas G. Brink}
\affiliation{Department of Astronomy, University of California, Berkeley, CA 94720-3411, USA}

\author{WeiKang Zheng}
\affiliation{Department of Astronomy, University of California, Berkeley, CA 94720-3411, USA}

\author[0000-0003-4610-1117]{Tam\'as Szalai}
\affiliation{Department of Experimental Physics, Institute of Physics, University of Szeged, H-6720 Szeged, D{\'o}m t{\'e}r 9, Hungary}
\affiliation{MTA-ELTE Lend\"ulet "Momentum" Milky Way Research Group, Szent Imre H. st. 112, 9700 Szombathely, Hungary}

\author{Joel Johansson}
\affiliation{Department of Physics, The Oskar Klein Center, Stockholm University, AlbaNova, 10691 Stockholm, Sweden} 

\author[0000-0002-4410-5387]{Armin Rest}
\affiliation{Space Telescope Science Institute, 3700 San Martin Drive, Baltimore, MD 21218, USA}
\affiliation{Department of Physics and Astronomy, Johns Hopkins University, 3400 North Charles Street, Baltimore, MD 21218, USA}

\author[0000-0001-9038-9950]{Schuyler D.~Van Dyk}
\affiliation{Caltech/IPAC, Mailcode 100-22, Pasadena, CA 91125, USA}

\author{Jennifer Andrews}
\affiliation{Gemini Observatory, 670 N. Aohoku Place, Hilo, Hawaii, 96720, USA}

\author[0000-0002-5221-7557]{Chris Ashall}
\affiliation{Department of Physics, Virginia Tech, Blacksburg, VA 24061, USA}
\affiliation{Institute for Astronomy, University of Hawai'i at Manoa, 2680 Woodlawn Dr., Hawai'i, HI 96822, USA}

\author[0000-0002-0141-7436]{Geoffrey C. Clayton}
\affiliation{Department of Physics \& Astronomy, Louisiana State University, Baton Rouge, LA 70803, USA}
\affiliation{Space Science Institute,
4765 Walnut St, Suite B
Boulder, CO 80301, USA}

\author[0000-0001-9419-6355]{Ilse De Looze}
\affiliation{Sterrenkundig Observatorium, Ghent University, Krijgslaan 281 - S9, B-9000 Gent, Belgium}

\author[0000-0002-7566-6080]{James M. DerKacy}
\affiliation{Space Telescope Science Institute, 3700 San Martin Drive, Baltimore, MD 21218, USA}

\author{Michael Dulude}
\affiliation{Space Telescope Science Institute, 3700 San Martin Drive, Baltimore, MD 21218, USA}

\author{Ryan J. Foley}
\affiliation{Department of Astronomy and Astrophysics, University of California, Santa Cruz, CA 95064, USA}

\author{Suvi Gezari}
\affiliation{Space Telescope Science Institute, 3700 San Martin Drive, Baltimore, MD 21218, USA}

\author[0000-0001-6395-6702]{Sebastian Gomez}
\affiliation{Space Telescope Science Institute, 3700 San Martin Drive, Baltimore, MD 21218, USA}

\author{Shireen Gonzaga}
\affiliation{Space Telescope Science Institute, 3700 San Martin Drive, Baltimore, MD 21218, USA}

\author{Siva Indukuri}
\affiliation{Department of Physics and Astronomy, Johns Hopkins University, 3400 North Charles Street, Baltimore, MD 21218, USA}

\author[0000-0001-5754-4007]{Jacob Jencson}
\affiliation{IPAC, Mail Code 100-22, Caltech, 1200 E.\ California Blvd., Pasadena, CA 91125}

\author{Mansi Kasliwal}
\affiliation{Cahill Center for Astrophysics, California Institute of Technology, 1200 E. California Blvd. Pasadena, CA 91125, USA}

\author[0009-0003-8380-4003]{Zachary G. Lane}
\affiliation{School of Physical and Chemical Sciences — Te Kura Matū, University of Canterbury, Private Bag 4800, Christchurch 8140, New Zealand}

\author{Ryan Lau}
\affiliation{NSF's NOIRLab, 950 N. Cherry Avenue, Tucson, 85719, AZ, USA}

\author{David Law}
\affiliation{Space Telescope Science Institute, 3700 San Martin Drive, Baltimore, MD 21218, USA}

\author[0000-0001-5788-5258]{Anthony Marston}
\affiliation{European Space Agency (ESA), ESAC, 28692 Villanueva de la Canada, Madrid, Spain}

\author[0000-0002-0763-3885]{Dan Milisavljevic}
\affiliation{Purdue University, Department of Physics and Astronomy, 525 Northwestern Ave, West Lafayette, IN 4790720, USA}

\author[0000-0002-2432-8946]{Richard O'Steen}
\affiliation{Space Telescope Science Institute, 3700 San Martin Drive, Baltimore, MD 21218, USA}

\author[0000-0002-2361-7201]{Justin Pierel}
\affiliation{Space Telescope Science Institute, 3700 San Martin Drive, Baltimore, MD 21218, USA}

\author[0000-0003-2445-3891]{Matthew Siebert}
\affiliation{Space Telescope Science Institute, 3700 San Martin Drive, Baltimore, MD 21218, USA}

\author{Michael Skrutskie}
\affiliation{Department of Astronomy, University of Virginia, Charlottesville, VA 22904-4325, USA}

\author[0000-0002-7756-4440]{Lou Strolger}
\affiliation{Space Telescope Science Institute, 3700 San Martin Drive, Baltimore, MD 21218, USA}

\author[0000-0002-1481-4676]{Samaporn Tinyanont}
\affiliation{National Astronomical Research Institute of Thailand (NARIT), Chiang Mai, 50180, Thailand}

\author{Qinan Wang}
\affiliation{Space Telescope Science Institute, 3700 San Martin Drive, Baltimore, MD 21218, USA}

\author{Brian Williams}
\affiliation{Observational Cosmology Lab, NASA Goddard Space Flight Center, Code 665, Greenbelt, MD 20771, USA}

\author{Lin Xiao}
\affiliation{Department of Physics, College of Physical Sciences and Technology, Hebei University, Baoding 071002, China}

\author[0000-0002-6535-8500]{Yi Yang}
\affiliation{Department of Astronomy, University of California, Berkeley, CA 94720-3411, USA}

\author[0000-0001-7473-4208]{Szanna Zs\'iros}
\affiliation{Department of Experimental Physics, Institute of Physics, University of Szeged, H-6720 Szeged, Hungary}

\correspondingauthor{Melissa Shahbandeh}
\email{mshahbandeh@stsci.edu}

\begin{abstract}
Dust from core-collapse supernovae (CCSNe), specifically Type~IIP SNe, has been suggested to be a significant source of the dust observed in high-redshift galaxies. 
CCSNe eject large amounts of newly formed heavy elements, which can condense into dust grains in the cooling ejecta. However, infrared (IR) observations of typical CCSNe generally measure dust masses that are too small to account for the dust production needed at high redshifts.
Type~IIn SNe, classified by their dense circumstellar medium (CSM), are also known to exhibit strong IR emission from warm dust, but the dust origin and heating mechanism have generally remained unconstrained because of limited observational capabilities in the mid-IR. 
Here, we present a {\it JWST}/MIRI Medium Resolution Spectrograph (MRS) spectrum of the Type IIn SN~2005ip nearly 17 years post-explosion. 
The Type~IIn SN~2005ip is one of the longest-lasting and most well-studied
SNe observed to date. 
Combined with a {\it Spitzer} mid-IR spectrum of SN~2005ip obtained in 2008, this data set provides a rare 15-year baseline, allowing for a unique investigation of the evolution of dust. 
The {\it JWST} spectrum shows a new high-mass dust component ($\gtrsim0.08$ \msolar) that is not present in the earlier {\it Spitzer} spectrum. 
Our analysis shows dust likely formed over the past 15 years in the cold, dense shell (CDS), between the forward and reverse shocks. 
There is also a smaller mass of carbonaceous dust ($\gtrsim0.005$ \msolar) in the ejecta. 
These observations provide new insights into the role of SN dust production, particularly within the CDS, and its potential contribution to the rapid dust enrichment of the early Universe.
\end{abstract}


\section{Introduction}
\label{sec:intro}

Core-collapse supernovae (CCSNe) mark the deaths of massive stars ($>$8 \msolar) triggered by the collapse of their iron cores. SNe IIP are the most common type of CCSN \citep{li11,smith11}, and for over 50 years they have been considered as a possible dominant source for the observed large amounts of dust observed at high redshifts \citep[$z>6$; e.g.,][]{cernuschi67,dwek07}. Such large amounts of dust in young galaxies underscore the need for rapid dust formation. CCSNe eject large amounts of heavy elements, such as carbon, oxygen, silicon, and iron, which can condense into dust grains within months to years post-explosion. The short Myr timescales of massive star SNe are much faster than the Gyr main-sequence lifetimes of lower-mass stars that have also been proposed as sources of dust \citep{dwek07,gall11}.

Theoretical models have succeeded in condensing sufficient amounts of dust (0.1--1 \msolar) in the expanding SN ejecta \citep{todini01,nozawa03,nozawa08}.
However, the dust yields inferred from Spitzer Space Telescope (hereafter \sst) observations in local CCSNe (mostly SNe~IIP) are often 2-3 orders of magnitude too small \citep[e.g.,][and those within]{zsiros24}. However, more recent observations enabled by larger telescopes and new technologies are beginning to challenge those results. For example, Galactic SN remnants (SNRs) and the very nearby ($\sim$50 kpc) SN 1987A reveal that massive cold dust reservoirs may be hiding significant dust quantities that, to be detected, require observations at longer mid-infrared (MIR) wavelengths at very late epochs \citep[e.g.,][]{matsuura11,temim17,delooze17,chawner20}. In fact, recent \jwst\ Mid-Infrared Instrument (MIRI) imaging of the Type IIP SN~2004et uncovered one of the largest newly formed ejecta dust masses in an extragalactic SN besides SN 1987A \citep{shahbandeh23}. 
 
Type IIn SNe (see \citealt{filippenko97,smith17a}~for reviews) have gained considerable attention over the past decade and have also become the focus of an increasing number of SN dust studies. Representing $<$10\% of all CCSN events \citep{li11,smith11}, these SNe stand out by the ``narrow'' ($\sim$100 \kms) lines formed from a high density and slow-moving circumstellar medium (CSM; \citealp{schlegel90,smith17a}). Compared to SNe IIP, SNe IIn tend to have an increased likelihood of exhibiting bright, late-time MIR emission from dust, which can have derived masses $\sim$1-2 orders of magnitude larger than SNe~IIP \citep[e.g.,][]{fox11,szalai19,szalai21}. While SNe~IIP are more frequent in the local universe \citep{li11}, Type~IIn progenitors are commonly linked to very massive stars, which may have been more prevalent during the Epoch of Reionization given an expected top-heavy initial mass function (IMF; \citealp{chary08,dave08}), and therefore could have contributed more cosmic dust at early times \citep{cherchneff10}. 

The origin of the SN IIn dust is not well-constrained. The evolving blue shift in the optical spectral line profiles is consistent with continuous dust growth in the ejecta or the cold, dense shell (CDS) that forms behind the forward shock \citep[e.g.,][]{smith08c,smith09,smith12,gall14,bevan19,bevan20,smith20}. Alternatively, geometries derived from models of the MIR spectral energy distribution (SED) place the majority of dust at radii beyond the likely forward shock radius, consistent with pre-existing dust in the large, dense CSM \citep[e.g.][]{fox11}. Of course, contributions to the dust emission may arise from multiple components. Unlike SNe~IIP, the dust in either scenario can be continuously heated to relatively high temperatures for years, if not decades, post-explosion by radiation arising from the ongoing shock interaction \citep{fox13a,vandyk13}. 

The Type IIn SN 2005ip, discovered in NGC~2906 (d $\approx$ 30~Mpc) on November 5, 2005, \citep{boles05}, is one of the most well-observed SNe IIn during the Cold and Warm \sst~eras \citep{fox09,smith09,fox10,stritzinger12,katsuda14,smith17,baknielsen18,bevan19,szalai19,fox20,szalai21}.
\citet{fox09} first reported a 3-year IR light-curve plateau indicative of warm dust. \citet{smith09} analyzed the spectral evolution of \ip, showing blue-shifted line profiles indicative of new dust formation in both post-shock CDS (intermediate-width components) and SN ejecta (broad emission lines). These appeared 100 days after the explosion, well before the onset of the late-time CSM interaction plateau. 
The relative contributions were not quantified, but \citet{bevan19} go on to model the evolution of the optical spectral lines with DAMOCLES. 
They find that the observed line asymmetries require the formation of nearly 0.1~\msolar over the first ten years. 

A \sst/IRS \citep{houck04} MIR spectrum (5$-$14~$\micron$) of the SN at $\sim$3 years post-explosion was obtained just before the \sst~Warm Mission (\citealt{fox10}). The spectrum is featureless and can be fit with two dust components. \citet{fox10} conclude the hotter ($\sim$900~K), less massive component ($\sim10^{-4}-10^{-3}$~\msolar) is likely dominated by the newly formed dust already discussed above. In fact, \citet{bevan19} estimate the amount of dust formed in SN 2005ip at $\sim$3 years is only $\sim10^{-3}$~\msolar~(their figure 11). 
Although the \citet{bevan19} models are found to be consistent with dust formation in the ejecta, it is essential to note that this does not mean the models are {\it inconsistent} with dust formation in the CDS. We reiterate that the predominant conclusion from the other papers is that most dust must have formed in the CDS, as this is where the bulk of the asymmetries are observed.

A second, cooler ($\sim$500 K), more massive component ($\sim10^{-2}$~\msolar) is also observed in the {\it Spitzer}/IRS spectrum, which \citet{fox10} attribute to an additional dust component with a geometry consisting of a spherical shell at or beyond the forward shock. They conclude that this colder, more massive dust component has both physical and geometric properties that are most consistent with a pre-existing CSM, although there were admittedly large uncertainties. The emission resulting from the ongoing CSM interaction was considered the primary heating source for any dust. 

Making definitive statements about the thermal-IR dust emission in \ip\ has been challenging because of the lack of sensitive MIR instrumentation since the end of the Cold Spitzer mission. Late-time MIR observations ($>$ 4.5 \micron) offer the opportunity to probe cooler dust in extragalactic SNe, providing new opportunities for detailed dust characterization in SN 2005ip for the first time since the {\sst}~Warm Mission began in 2009. Here, we present {\it JWST} MIRI spectroscopy of SN~2005ip. 
In Section~\ref{sec:obs}, we present the observations and reduction techniques. 
Section~\ref{sec:modeling} describes our spectral fitting and dust modeling procedures.
Section~\ref{sec:scenarios} explores the origin and heating mechanism of dust in \ip. 
Section~\ref{sec:conclusions} presents a summary and conclusions.

\section{Observations and Data Reduction} \label{sec:obs}
\subsection{\jwst/MIRI Medium Resolution Spectroscopy} \label{subsec:MRS}
We obtained a single epoch of SN~2005ip with \jwst/MIRI's Medium Resolution Spectrometer (MRS; \citealt{wells15, argyriou23, bushouse22}) on UT 2023 April 20 (JD 2456402), 6376 days post-discovery (UT dates are used throughout this paper). 
The data consist of $R$ = $\lambda$/$\Delta\lambda$ = 1,330-3,750 spectroscopy spanning 4.9 to 27.9 \micron, including all sub-bands, to measure the expected dust continuum. 
Table~\ref{tab:tab1} provides a summary of the observation parameters. 
These observations are part of PID 1860 (PI: Fox), which aims to probe the dust properties around SNe~IIn.

\subsubsection{Creating the Unsubtracted Data Cube}
Uncalibrated data were initially downloaded from MAST\footnote{\dataset[DOI:10.17909/q96n-2296]{\doi{10.17909/q96n-2296}}}. All data were processed with JWST calibration pipeline version 1.14.0 \citep{bushouse22}, using CRDS pmap 1263. The global thermal background is roughly 10 to 40 times larger than the SN flux. This makes estimating the local background both critical and challenging. Even small variability in the background across the field of view (FOV) can have a potentially large impact on our source spectrum. 
This is particularly true at the longest wavelengths, where the thermal background tends to be the largest, and the signal-to-noise from the source tends to be the weakest due to a combination of lower signal and the known reduced count rate\footnote{\url{https://www.stsci.edu/contents/news/jwst/2023/miri-mrs-reduced-count-rate-update}}.

\begin{deluxetable}{ l l l l l l}
\setlength{\tabcolsep}{1pt}
\small
\caption{{\it JWST}/MRS Spectroscopy of SN 2005ip \label{tab:tab1}}
\tablehead{
\colhead{Ch} & \colhead{FOV} & \colhead{Sub-Band} & \colhead{Wavelengths} & \colhead{Resolving Power} & \colhead{Integration} \\
\colhead{}     &    \colhead{(\arcsec)} & \colhead{} & \colhead{(\micron)} & \colhead{($\lambda/\Delta\lambda$)} & \colhead{(sec)}
    }
\startdata
\hline
\multirow{3}{*}{1} & \multirow{3}{*}{3.2$\times$3.7} & Short (A)  & 4.90–5.74 & 3,320–3,710 & 1337\\
                   &                                  & Medium (B) & 5.66–6.63 & 3,190–3,750 & 1337\\
                   &                                  & Long (C)   & 6.53–7.65 & 3,100–3,610 & 1337\\
\hline
\multirow{3}{*}{2} & \multirow{3}{*}{4.0$\times$4.8} & Short (A)  & 7.51–8.77 & 2,990–3,110 & 1337\\
                   &                                  & Medium (B) & 8.67–10.13 & 2,750–3,170 & 1337\\
                   &                                  & Long (C)   & 10.02–11.70 & 2,860–3,300  & 1337\\
\hline
\multirow{3}{*}{3} & \multirow{3}{*}{5.2$\times$6.2} & Short (A)  & 11.55–13.47 & 2,530–2,880 & 1337\\
                   &                                  & Medium (B) & 13.34–15.57 & 1,790–2,640 & 1337\\
                   &                                  & Long (C)   & 15.41–17.98 & 1,980–2,790  & 1337\\
\hline
\multirow{3}{*}{4} & \multirow{3}{*}{6.6$\times$7.7} & Short (A)  & 17.70–20.95 & 1,460–1,930 & 1337\\
                   &                                  & Medium (B) & 20.69–24.48 & 1,680–1,770 & 1337\\
                   &                                  & Long (C)   & 24.19–27.90 & 1,630–1,330  & 1337\\
\hline
\enddata
\end{deluxetable}

\begin{figure}
    \centering
    \includegraphics[width=0.45\textwidth]{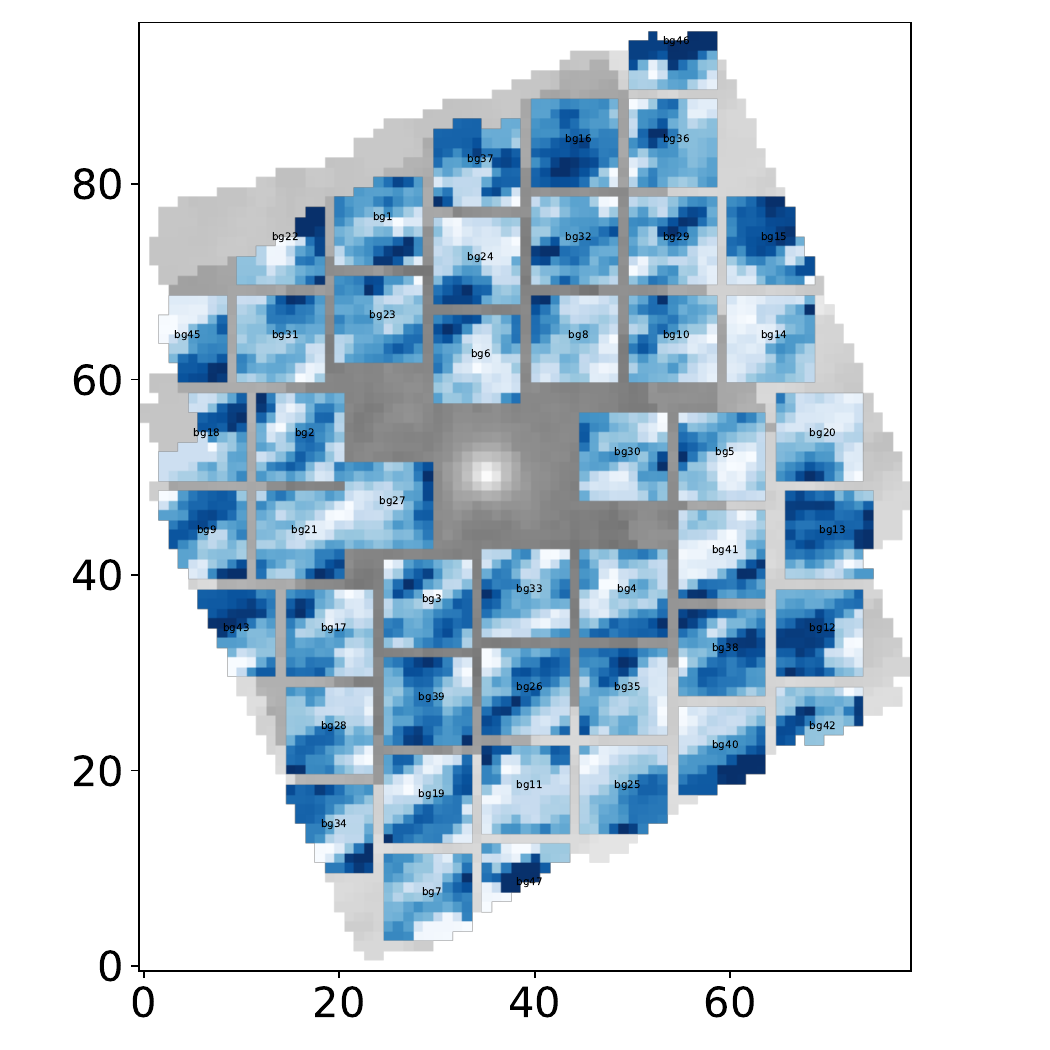}
    \caption{A snapshot of the \ip\ cube with the sampled background regions over-plotted in blue shades.}
    \label{fig:bgs_cube}
\end{figure}

We explored several techniques for estimating the local background, including implementing a background annulus during the extraction, averaging an array of apertures around the FOV, and performing a 2D interpolation and estimate of the local background. We ultimately implemented the following technique that averages the background at different locations around the SN as detailed below.

\subsubsection{Constructing and Subtracting the Master Background}
To address the thermal background challenge for this paper, we took the following steps:
\paragraph{Background Estimation} Specific regions of the data cube, identified as background, are manually selected. 
Spectra from these regions are extracted and averaged to construct a master background. This is similar to procedures followed for other MRS spectra of SNe \citep{shahbadneh24,derkacy24,ashall24}.
We generate a median background based on sampling 48--62 different positions across the cube FOV (Figure~\ref{fig:bgs_cube}). 

\begin{figure*}[!ht]
    \centering
    \includegraphics[width=0.95\textwidth]{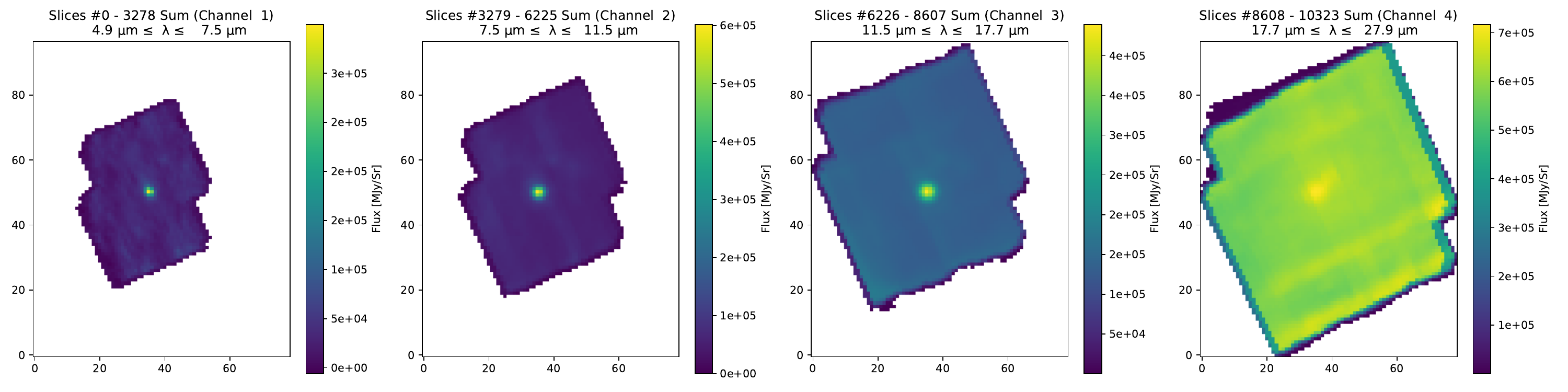}
     \includegraphics[width=0.95\textwidth]{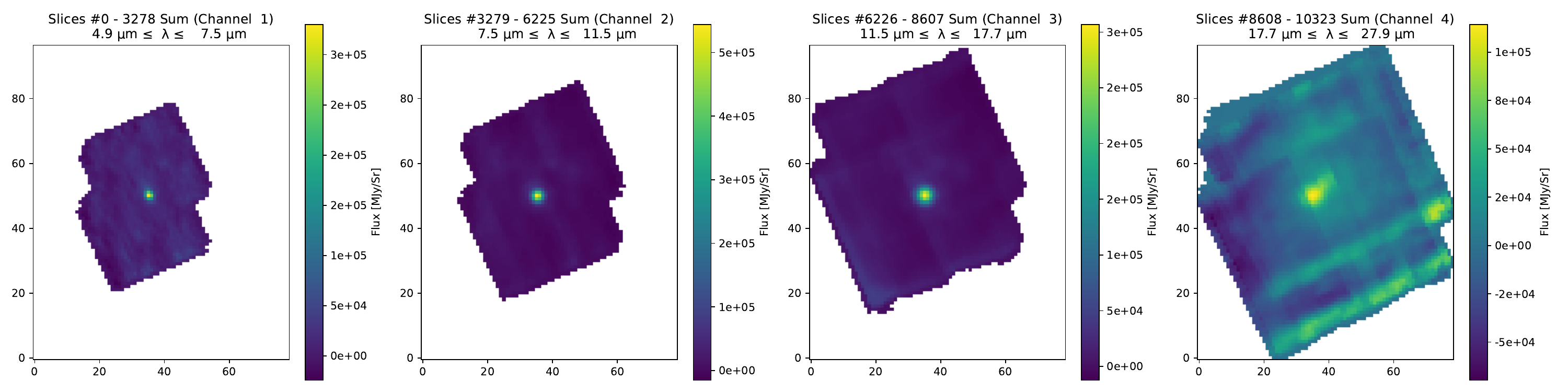}
    \caption{\textit{Top:} MIRI/MRS cube of SN~2005ip before background subtraction for each of the 4 MRS channels. Each channel is a collapsed sum of all its slices.
    \textit{Bottom:} Same as top, but after background subtraction.
    }
    \label{fig:cubeimage}
    \includegraphics[width=0.95\textwidth]{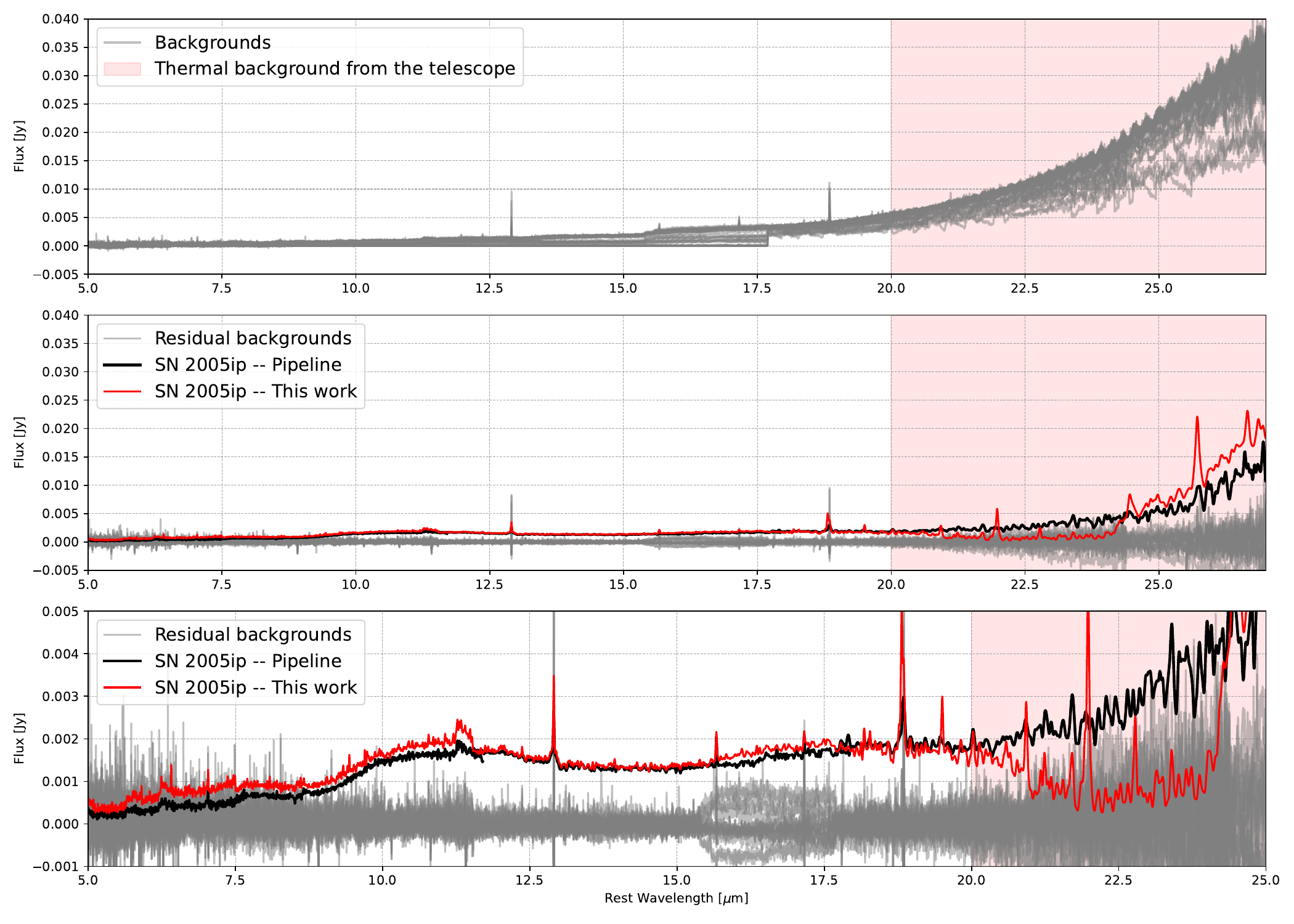}
    \caption{The extracted spectra (grey) in the MRS cube from all the selected regions in Figure \ref{fig:bgs_cube}. Overplotted are the extractions from the pipeline (black) and this work (red). \textit{Top:} Before background subtraction. \textit{Middle:} After background subtraction. \textit{Bottom:} Same as the middle panel, but scaled to better show the features of SN 2005ip.}
    \label{fig:bgs_specs}
\end{figure*}

\paragraph{Background Subtraction} The median background is subtracted from each spaxel in the data cube, producing a background-subtracted cube. This process is essential to minimize the impact of background variations on the final spectrum. The notebook is available on Github\footnote{\url{https://github.com/shahbandeh/MIRI_MRS}}.
Figure~\ref{fig:cubeimage} shows the collapsed cube in each channel before and after background subtraction.
Similarly, Figure \ref{fig:bgs_specs} shows the extractions for each spaxel before and after background subtraction. 

\subsubsection{Analysis of Residual Backgrounds}
After background subtraction, we perform additional analysis to ensure minimal residual background:
\paragraph{Residual Extraction} Residual background spectra are extracted from the subtracted cube and averaged to evaluate the effectiveness of the background subtraction.

\paragraph{Comparison} The residual spectra are compared to the original background spectra to identify any remaining background contamination and assess the noise reduction achieved.

\subsubsection{Final spectrum extraction}
\begin{figure*}[!ht]
    \centering
    \includegraphics[width=0.9\textwidth]{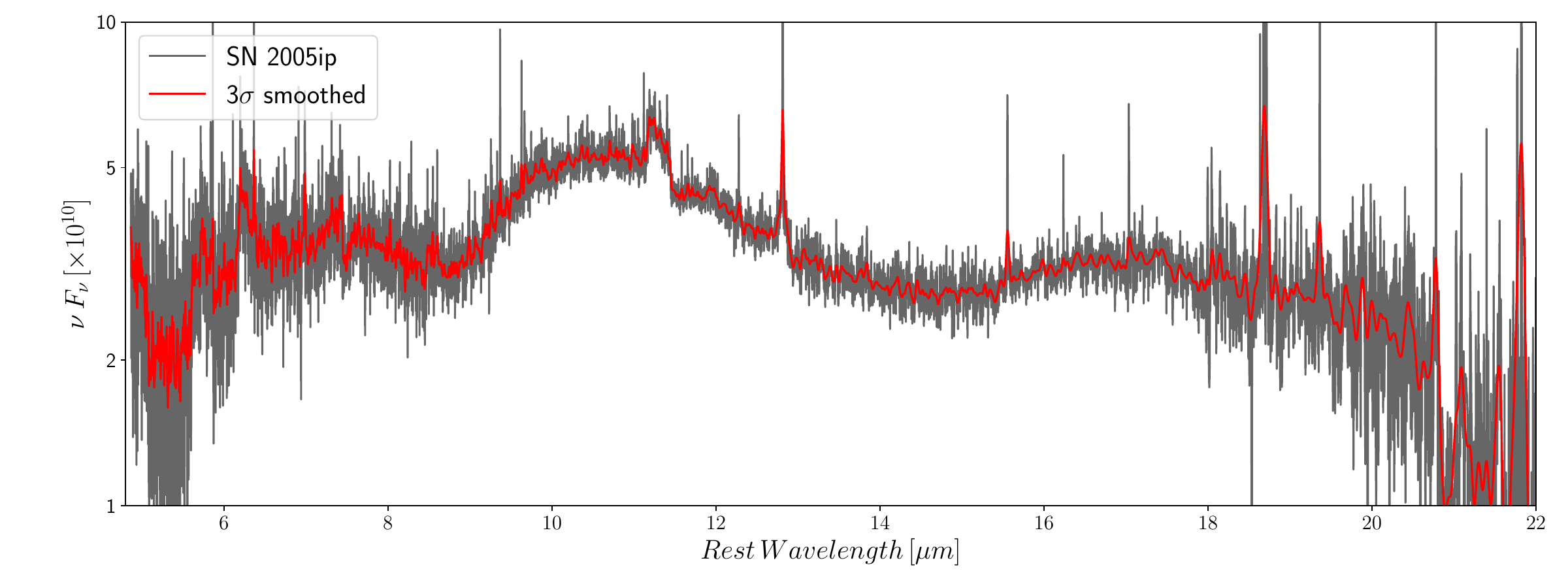}
    \caption{Final extraction of MIRI/MRS spectrum of \ip\ obtained on April 20, 2023, at 6376~days post discovery.}
    \label{fig:jw_spec}
    \centering
    \includegraphics[width=0.8\textwidth]{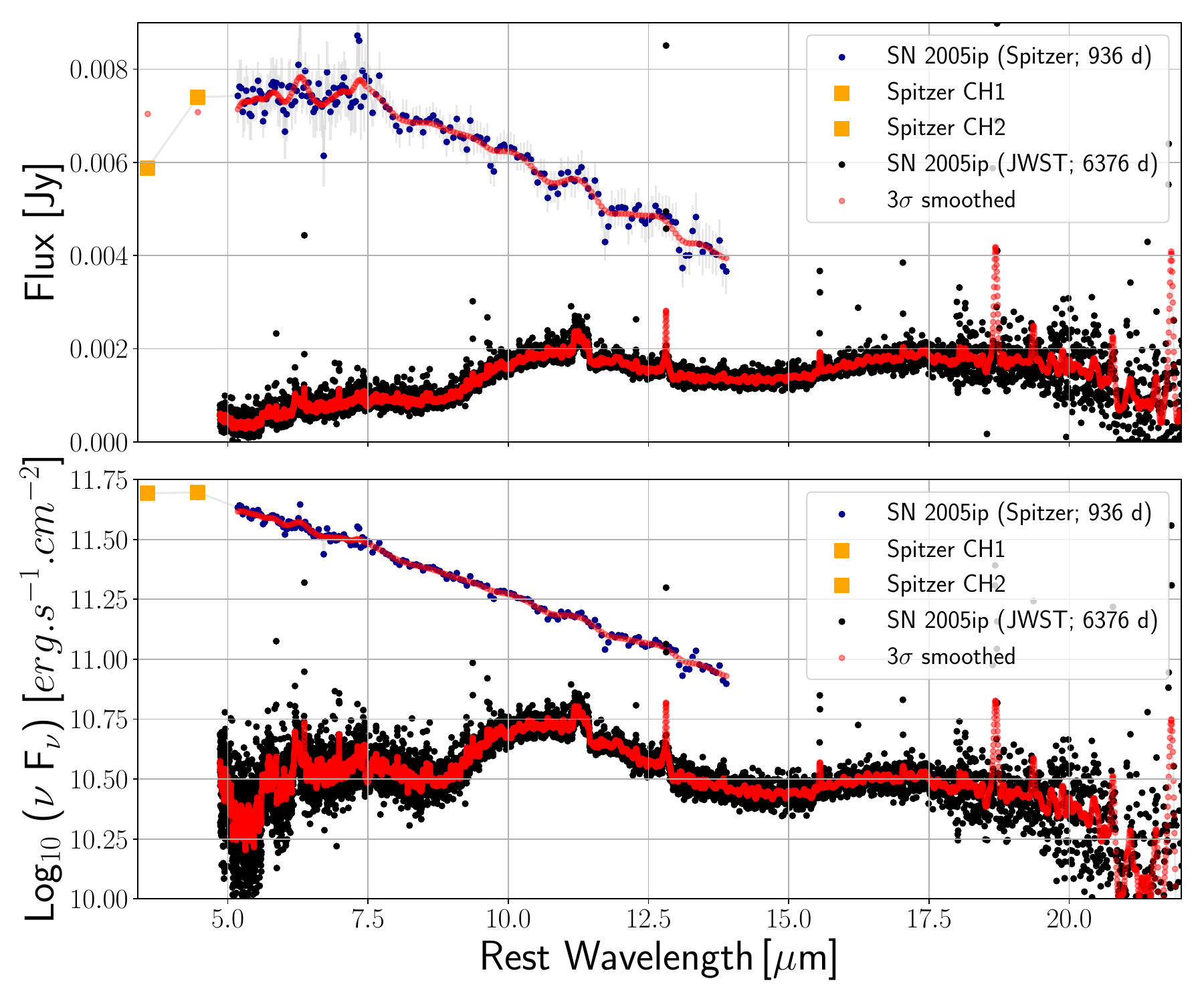}
    \caption{Direct comparison of the {\it Spitzer} and {\it JWST} MIR spectra of SN 2005ip, obtained roughly 15 years apart. Both spectra are plotted on the same scale for comparison purposes.
    }
    \label{fig:spectra}
\end{figure*}

The final step involves extracting the spectrum of the target source performed at the position of the SN using the \textsc{Extract1dStep} command in the \jwst\ calibration pipeline. 
Before performing this extraction, we first searched for the brightest source in the cube, which we identified as the SN, rather than relying solely on the coordinates provided in the header.
While we mostly used default parameters, a full list can be found in the Github notebook.
This extraction step uses a circular aperture that varies in radius with wavelength to account for the wavelength-dependent point spread function (PSF) of the instrument. 
The extraction-related vectors are found in the Advanced Scientific Data Format (asdf) \texttt{extract1d} reference file version 0004. 
The extracted spectrum represents the final scientific product, with background effects minimized to the extent possible.
No additional background subtraction was used, as the background estimation technique is considered to have adequately removed the local background at the position of the source (see Figures~\ref{fig:cubeimage} and \ref{fig:bgs_specs}). 
Figure~\ref{fig:jw_spec} plots the final MRS extraction of \ip\ and Figure~\ref{fig:spectra} compares the {\it JWST} and {\it Spitzer} spectra, obtained roughly 15 years apart.

An interesting spectrum feature is the drop in flux at $>$17 \micron, where Channel 4 begins. 
As discussed further in Section \ref{sec:modeling}, we chose not to fit these regions ($>$22 \micron) due to the challenges of background subtraction at longer wavelengths, where the thermal background dominates. 
These wavelengths are crucial for differentiating between models and strongly constrain the cold dust mass. 
Additionally, Channel 4 ($>$17.8 \micron) in the MRS is known to experience a reduced count rate over time, and we have applied a correction factor to account for this. 
As a result, we place less emphasis on these wavelengths in our modeling.

\begin{figure*}
    \centering
    \includegraphics[width=0.69\textwidth]{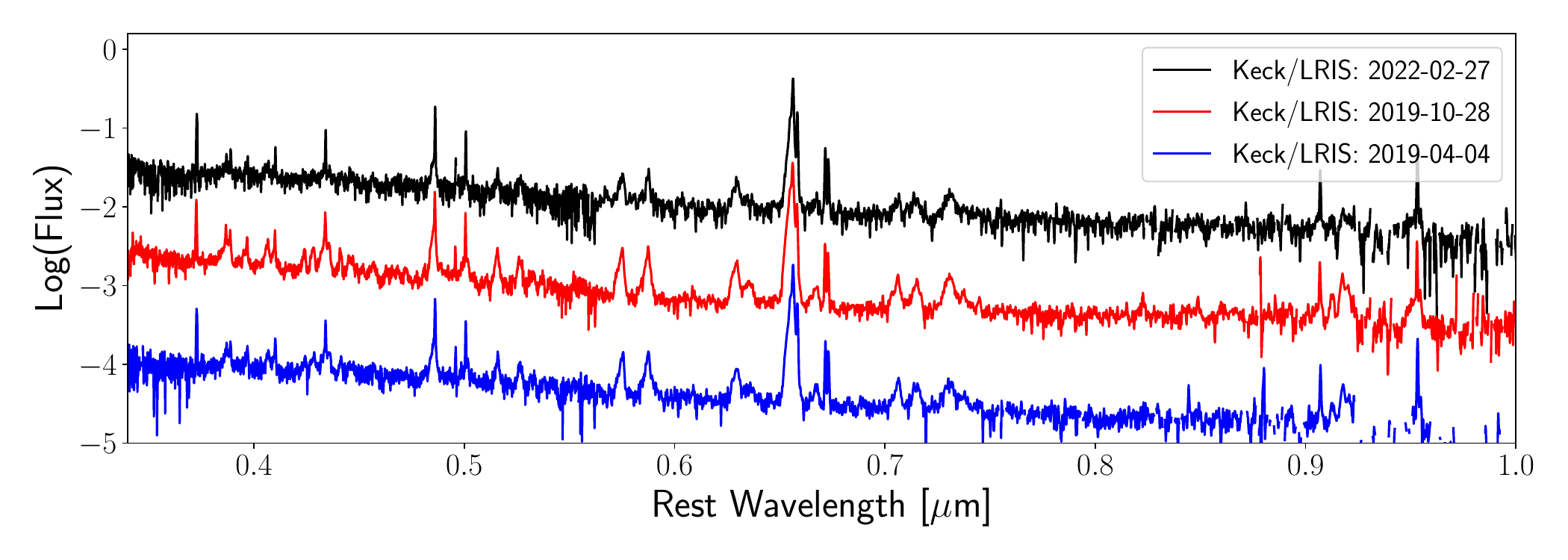}
    \includegraphics[width=0.28\textwidth]{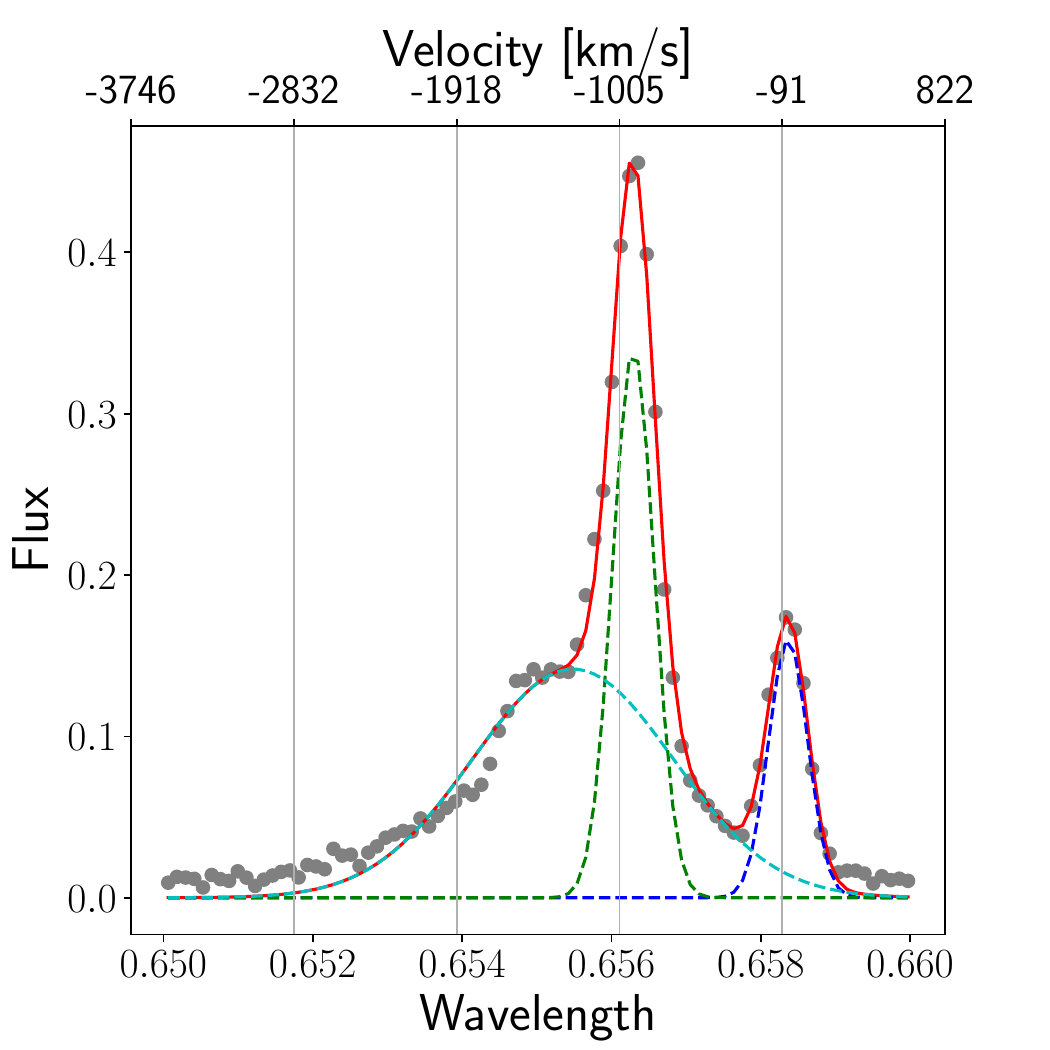}
    \caption{Late-time optical spectrum of SN~2005ip obtained with the Keck/LRIS. The presence of the H$\alpha$ emission line signifies ongoing shock interaction with pre-existing CSM.}
    \label{fig:optical}
\end{figure*}

\subsection{Optical Spectroscopy}\label{subsec:optical}
Late-time spectra of \ip\ were obtained with the Keck Low-Resolution Imaging Spectrometer (LRIS; \citealp{oke95}) shown in Figure~\ref{fig:optical}. 
The spectra were acquired with the slit oriented at or near the parallactic angle to minimize slit losses caused by atmospheric dispersion \citep{filippenko82}. The LRIS observations utilized the $1\arcsec$ slit, 600/4000 grism, and 400/8500 grating. 
Data reduction followed standard techniques for CCD processing and spectrum extraction \citep{silverman12}
utilizing IRAF
\citep{tody86} routines and custom Python and IDL codes.\footnote{\url{https://github.com/ishivvers/TheKastShiv}}
The most recent LRIS spectra (from 2017 and later) were processed using the LPipe data-reduction pipeline \citep{perley19}. 
Low-order polynomial fits to comparison-lamp spectra were used to calibrate the wavelength scale, and small adjustments derived from night-sky lines in the target frames were applied. 
The spectra were flux calibrated using appropriate spectrophotometric standard stars observed on the same night, at similar airmasses, and with an identical instrument configuration.

\section{Modeling the MIR data} \label{sec:modeling}
Here, we describe the modeling of both MIR spectra. 
Throughout this work, we assume that the MIR flux is dominated by thermal emission from dust. We also consider additional flux contributions, such as synchrotron emission, to ensure that most physical scenarios are properly investigated.

\subsection{Necessary Equations}\label{subsec:modelmath}
We follow similar procedures outlined in \citet[][also see \citealt{fox10, dwek19}]{shahbandeh23}. 
Since we do not know the precise origin of the dust, we must first consider different geometries.

\subsubsection{Dust sphere}\label{subsec:sphere}
A homogeneously expanding spherical geometry is commonly assumed in many papers discussing new dust formation in SNe.
The observed dust mass, $M_{\rm d}$, is most accurately written by taking into account not only the observed flux, $F_{\rm obs}$, but also the escape probability of the infrared photons from the emitting region, $P_{\rm esc}$, which can vary as a function of optical depth, $\tau$:
\begin{equation}\label{eqn:main} 
M_{\rm d} = {F_{\rm obs}(\lambda)\ d^2\ \over B(\lambda, T_{\rm d})\, \kappa(\lambda)P_{\rm esc}(\tau)},
\end{equation}
where
$\lambda$~is wavelength, $d$ the distance to the source, $\kappa(\lambda)$ is the dust mass absorption coefficient,
$B(\lambda, T_{\rm d})$ is the Planck function at wavelength $\lambda$, and $T_{\rm d}$ is the dust temperature. 
When $P_{\rm esc} \approx 1$, the dust is optically thin, and all the dust is accounted for. 
For $P_{\rm esc} < 1$, the dust is optically thick, and a large fraction of the dust may be undetected and, therefore, unaccounted for.
For a sphere only, $P_{\rm esc}(\tau)$ can be written as \citep{fox10, dwek19}:
\begin{equation}\label{eqn:Pesc}
P_{\rm esc}(\tau) = {3\over 4 \tau}\ \left[1-{1\over2\tau^2}+({1\over\tau}+{1\over2\tau^2})e^{-2\tau}\right], \, 
\end{equation}
where 





\begin{equation}
\label{eqn:tau}
\tau = \frac{3M_{\text{d}}(R_2 - R_1) \kappa(\lambda)}{4\pi(R_2^3 - R_1^3)}, \,
\end{equation}
and R$_2$~ and R$_1$~are the outer and inner radii of the dust shell, respectively. In the case of the sphere, R$_1$=0. In the case of low optical depth ($\tau<<1$), $P_{\rm esc}(\tau) \approx 1$. In the case of high optical depth ($\tau>>1$), $P_{\rm esc}(\tau) \approx \frac{3}{4\tau}$.

\subsubsection{Dust shell}\label{subsec:shell}
Some physical scenarios, such as the CDS, require a different geometry consisting of a shell of dust (as opposed to a sphere).
While the equation for $P_{\rm esc}$ does vary with geometry, it does not impact the overall model significantly. 
The spherical models in Equations~\ref{eqn:main} - \ref{eqn:tau} are still robust estimates for the dust mass and temperature and are preferred over simpler models that assume an optically thin scenario and do not include any estimate of $P_{\rm esc}$. 
Alternatively, the radii R$_1$~and R$_2$~may vary significantly. 
This is because the escape probability in Equation \ref{eqn:Pesc} was derived for a sphere. If all the dust from the spherical geometry is moved into a shell, the shell must have a larger outer (and inner) radius to spread out the dust grains and create the same optical depth. 
We discuss our calculations of radius and the values we choose for our final interpretation of each model in more detail below.

\subsubsection{Blackbody Radius}\label{sec:bb}
While Equation \ref{eqn:main} does fit for a radius, R$_2$, in a spherical, optically thick geometry, R$_2$~is unconstrained in optically thin cases, where $\tau \ll 1$. For such a scenario, an alternative approach to calculating the radius is to use the blackbody radius, R$_{\rm bb}$. 
The blackbody radius assumes a $\tau=1$~and can be interpreted as the minimum radius of a spherically symmetric shell. 
In other words, the dust may reside in a spherical shell with an even larger radius or may not even reside in a shell at all. 
However, if the radius of the shell were any smaller, the dust would become optically thick. 
The corresponding equation assumes that the observed flux emanates from a perfect blackbody with a $\tau\approx1$:
\begin{equation}
\label{equ:rbb}  
R_{\rm bb}(t)=\sqrt{\frac{L_{\rm obs}}{4 \pi \sigma T_{\rm d}^4}}=\sqrt{\frac{4 \pi d^2 \int F_{\text {obs }}(\lambda) d \lambda}{4 \pi \sigma T_{\rm d}^4}},
\end{equation}
where $L_{\rm obs}$~is the total integrated MIR luminosity measured empirically directly from the data.
The dust temperature, T$_{\rm d}$, used in this equation is the best fitting value derived from modeling the data using Equation \ref{eqn:main}. While the dust does not radiate as a perfect blackbody, the temperature remains a robust estimate in Equation \ref{equ:rbb}.

\subsection{Spectral Fitting of MIR Spectra}\label{subsec:fitting}
We fit both IR spectra (\sst~and \jwst) with Equation \ref{eqn:main} using similar methods as described in \citet{shahbandeh23}. 
We implement least-squares minimization using the Python \texttt{lmfit} package. 
We consider multiple dust components varying in temperature, mass, optical depth, and composition, where the composition is driven by the absorption coefficient, $\kappa$. 
We also consider various dust compositions, including O-rich dust species, such as Mg-silicates, and C-rich dust species, such as amorphous carbon or graphite \citep{wesson21, ercolano07, sarangi13}. 
The absorption and emission characteristics of these grains are derived from \cite{draine07} for silicates and polycyclic aromatic hydrocarbons (PAHs), and \cite{zubko04} for amorphous carbon (refer to \cite{draine07} or \cite{sarangi22} for absorption coefficients $\kappa$ values). 
We also include components for narrow line emission using Drude models\footnote{\url{https://docs.astropy.org/en/latest/modeling/physical\_models.html\#drude1d}} and synchrotron emission using power laws.
 
Qualitatively, even before fitting any models, the emissions in the \jwst spectrum suggest a cooler, massive Mg-silicate component (Cool Dust: CD) and a warmer amorphous carbon component (Warm Dust: WD). 
The dominant silicate features drive the CD component, while the WD component is suggested by the shorter wavelength, featureless emission that is apparent between 5-8~\micron. 
An amorphous carbon component is the most likely model capable of fitting this short-wavelength behavior.
We considered other models, including a power law representing synchrotron radiation, but none succeeded. 
There are also noticeable PAH features that have been incorporated into the model.
\begin{figure*}
    \centering
    \includegraphics[width=0.98\textwidth]{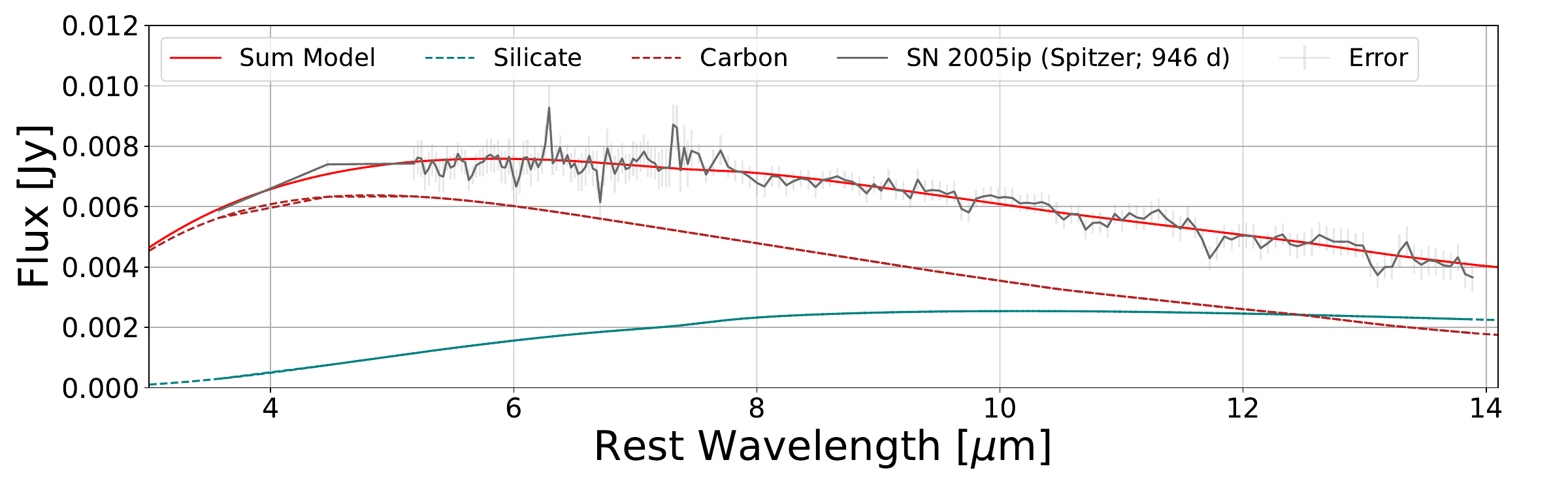}
    \caption{The observed MIR spectrum of SN 2005ip obtained with \sst\ at day 946. Overplotted are the best-fitting components with parameters listed in Table \ref{tab:spitzer_fit}.}
    \label{fig:sstmodel}
\end{figure*}

\begin{table}
\centering
\caption {\sst~Model Parameters \label{tab:spitzer_fit}}
\resizebox{\columnwidth}{!}{
\begin{tabular}{|l|c|c|}
\hline
\textbf{Component} & \textbf{Parameters} & \textbf{Best Fit}\tablenotemark{*} \\
\hline
\multirow{3}{*}{Mg-Silicate (CD)} & 
T [K] & 500$^{+160}_{-90}$ \\
& M$_{\rm d}$[\msolar] & 0.07$^{+\infty}_{-0.04}$ \\
& $R_{\rm d}[10^{16} \rm cm$] & 6.3$^{+1.5}_{-1.2}$ \\
\hline
\multirow{3}{*}{Amorphous Carbon (WD)} &
T [K] & 1060$^{+240}_{-90}$\\
& M$_{\rm d}$[\msolar] & 0.005$^{+0.004}_{-0.003}$ \\
& $R_{\rm d}[10^{16} \rm cm$] & 3.3$^{+0.6}_{-1.3}$ \\
\hline
\end{tabular}
}
\tablenotetext{*}{Values within the uncertainties will encompass the true value 95\% of the time.}
\end{table}

\begin{figure*}
    \includegraphics[width=0.95\textwidth]{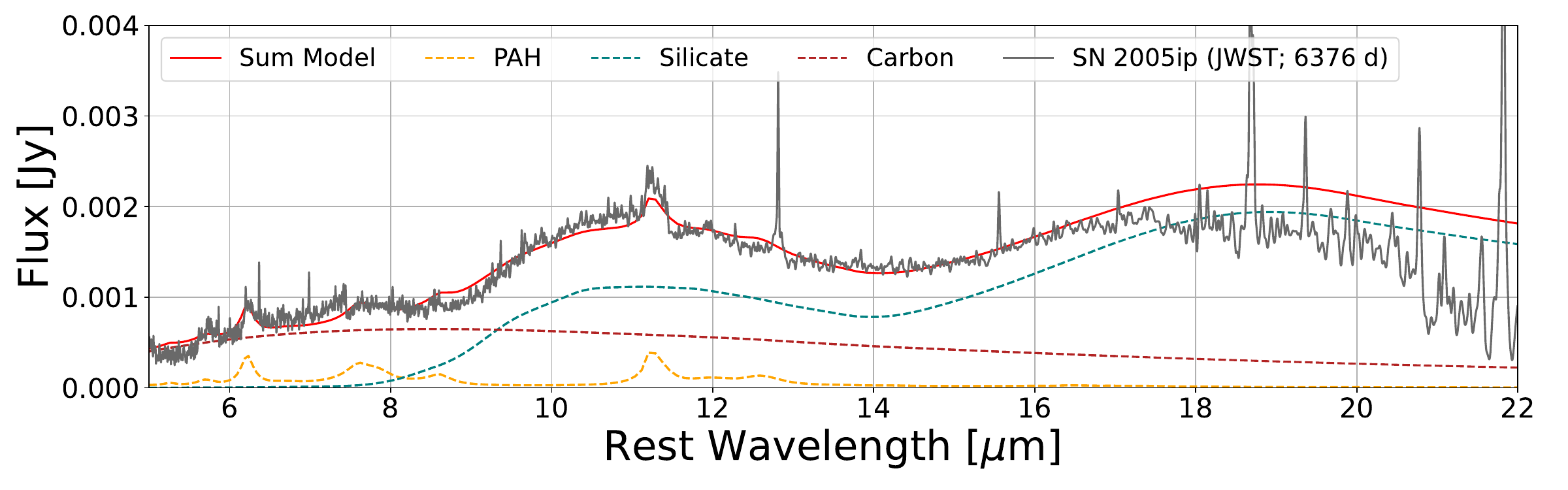}
    \caption{The observed MIR spectrum of \ip\ obtained with \jwst~at day 6376. Overplotted are the best-fitting components with parameters listed in Table \ref{tab:jwst_fit}.}
    \label{fig:jwstmodel}
\end{figure*}

\begin{table}
\centering
\caption{\jwst~Model Parameters \label{tab:jwst_fit}}
\resizebox{\columnwidth}{!}{
\begin{tabular}{|l|c|c|}
\hline
\textbf{Component} & \textbf{Parameters} & \textbf{Best Fit}\tablenotemark{*} \\
\hline
\multirow{3}{*}{Mg-Silicate (CD)} &
T [K] & 192$^{+1}_{-2}$ \\
& M$_{\rm d}$[\msolar] & 0.081$^{+0.004}_{-0.004}$ \\
& R$_{\rm d}[10^{18} \rm cm$] & 8.5$^{+\infty}_{-2.5}$\\
\hline
\multirow{3}{*}{Amorphous Carbon (WD)} &
T [K] & 580$^{+10}_{-10}$ \\
& M$_{\rm d}$[\msolar]  &  0.006$^{+0.001}_{-0.001}$ \\
& R$_{\rm d}[10^{16} \rm cm$] & 2.55$^{+0.09}_{-0.08}$ \\
\hline
\end{tabular}
}
\tablenotetext{*}{Values within the uncertainties will encompass the true value 95\% of the time.}
\end{table}

No obvious silicate features are present in the \sst~spectrum. The lack of these features may be due to the fact that the silicate dust component is not yet present or that the silicate features are suppressed by an optically thick geometry \citep{dwek15}. In either case, we assume the same two dust components (that is, a CD and WD component) for the model we use for the \sst~fit, although we allow the specific parameters of those components to vary. 
We considered many different scenarios in our analysis but only report our best-fitting results here.

Connecting the evolution of the spectra is a unique opportunity for \ip, as no other SN~IIn has multiple MIR spectra, particularly over such a long time. 
Using the same components in each model simplifies any interpretation of such evolution, which would be nearly impossible to constrain otherwise.
However, the assumption is reasonable given that \ip\ is known to be a relatively constant, slowly changing transient \citep[e.g.,][and those within]{smith17,fox20}. 
The MIR photometric evolution observed with \sst~at 3.6 and 4.5 \micron, in particular, shows a slow and steady decline \citep{szalai19,szalai21}, suggestive of no obvious newly appearing or rapidly disappearing component.

\begin{figure}
    \centering
    \includegraphics[width=0.85\columnwidth]{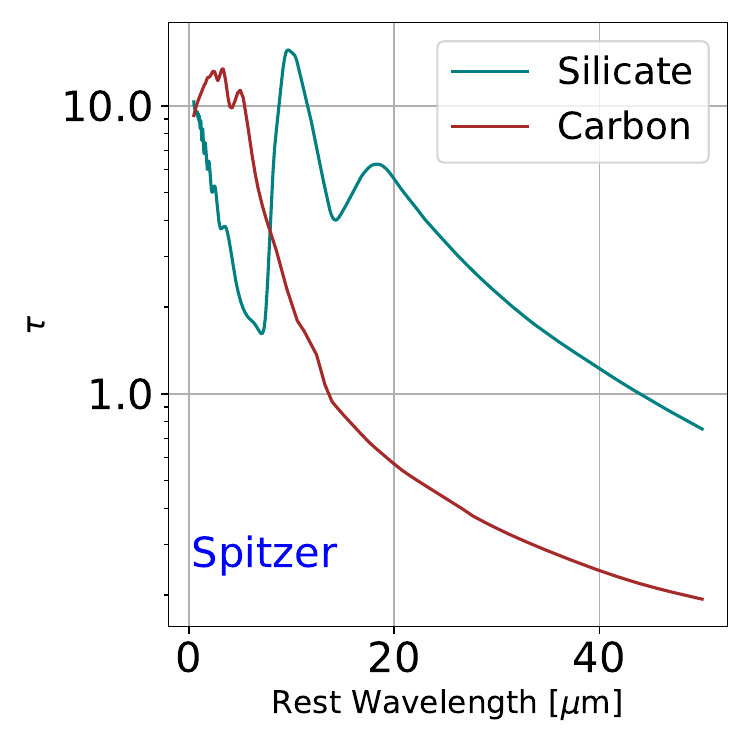}
    \vspace{1em}
    \includegraphics[width=0.85\columnwidth]{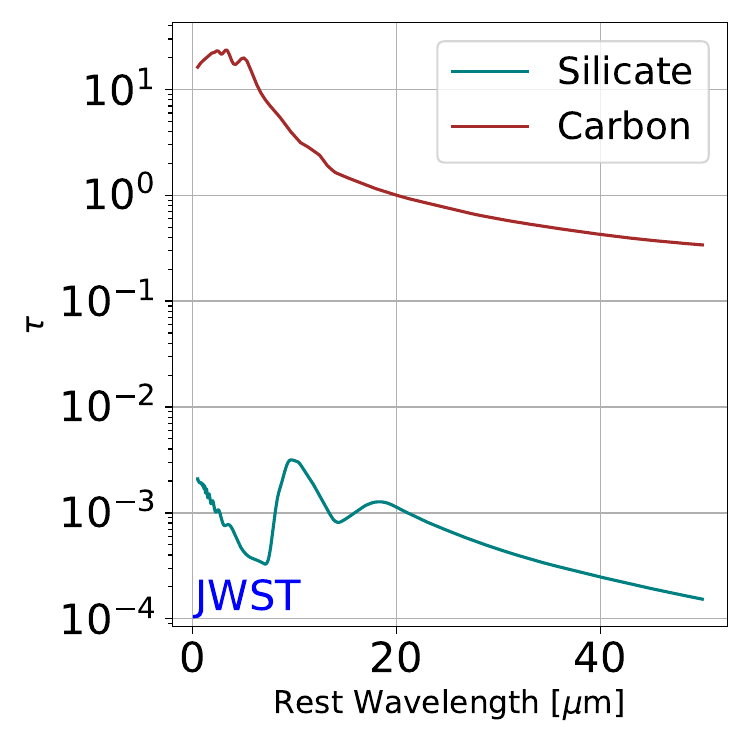}
    \caption{The optical depth, $\tau$, as a function of wavelength for the \sst~and \jwst~best-fitting models in Figures \ref{fig:sstmodel} and \ref{fig:jwstmodel}. The optical depth is calculated with Equation \ref{eqn:tau} using the 95\% values in Table \ref{tab:spitzer_fit} and \ref{tab:jwst_fit}. The most notable takeaway is that the optical depth decreases dramatically with time for \ip.}
    \label{fig:tau}
\end{figure}

Figures~\ref{fig:sstmodel} and \ref{fig:jwstmodel} show our best-fitting model in red, and Tables \ref{tab:spitzer_fit} and \ref{tab:jwst_fit} list the best-fitting parameters. 
The best fits were determined by minimizing $\chi^2$. 
Our uncertainties, however, are asymmetric (i.e., non-gaussian). 
The standard error does not give a good estimate of the uncertainties in the values. 
When standard errors fail, confidence intervals provide a much more robust estimate of the uncertainty. 
The tables, therefore, list the best fit and corresponding 95\% confidence intervals. 
In other words, values within the uncertainties will encompass the true value of each parameter 95\% of the time. 
Note that some of the confidence intervals extend to infinity.
For the dust mass, this is because, at some point, the dust geometry becomes so optically thick that adding more dust does not increase the overall flux. 
In the case of the dust radius, an optically thin dust component has no upper bound. 
Figure \ref{fig:tau} also plots the corresponding values for $\tau$~as a function of wavelength associated with the best-fitting parameters (Equation \ref{eqn:tau}).

It should again be noted here that Equation \ref{eqn:Pesc} assumes that the dust density and emissivity are uniformly distributed in a spherically symmetric and homogeneous sphere. 
The biggest limitation in using this equation is that the geometry (i.e., a shell) and heating mechanism (radiation from the forward and/or reverse shock) of our environment may be much more complicated, potentially creating a temperature gradient in a shell. 
However, we consider Equations \ref{eqn:main} - \ref{eqn:tau} sufficient for this analysis. 
A more complete treatment is beyond the scope of this paper, but is being explored (see \citealt{dwek24a, dwek24b}) and can be incorporated into future work.

\section{Dust Origin and Heating Mechanism}\label{sec:scenarios}
The general SN environment can be complicated, with many different possible geometries, origins, and heating mechanisms for the dust.
The dust may be newly formed or pre-existing. 
If newly formed, the dust may condense in the expanding SN ejecta \citep{kozasa89, wooden93} or in the CDS of post-shocked gas lying in between the forward and reverse shocks \citep{smith08b, sarangi22a}. 
If pre-existing, the dust may have formed in a steady wind or during a short-duration pre-SN outburst. 
In any scenario, several heating mechanisms are possible, including a thermal light echo from the peak SN flash, radiative heating from circumstellar interaction, and/or collisional heating by hot gas in the reverse shock \citep[e.g.,][]{fox10}.
Throughout this section, it is our task to build a consistent scenario that considers both the spectra, both of the components, and other previous results. 
Geometry will play an important part in the interpretation, and we consider various scenarios.

Tables \ref{tab:spitzer_fit} and \ref{tab:jwst_fit} list both the best-fit and 95\% confidence intervals for all models. 
In general, we use the lower confidence 95\% as a lower bound/limit for our interpretation of all parameters since it best encompasses the fit uncertainties. 
For the dust radius, this choice is meaningful for the optically thick fits but not necessarily the optically thin case (see Section \ref{sec:bb}). 
Figure~\ref{fig:tau} shows that an optically thick model works for most of the fits except the \jwst~CD component (Mg-Silicate). 
In this case, we instead calculate the blackbody radius using Equation \ref{equ:rbb} with the total integrated flux in the silicate component and the lower bound of the associated dust temperature in Table \ref{tab:jwst_fit}, which results in R$_{bb}=$1.9$\times10^{17}$~cm.

\begin{figure}
    \centering
    \includegraphics[width=0.49\textwidth]{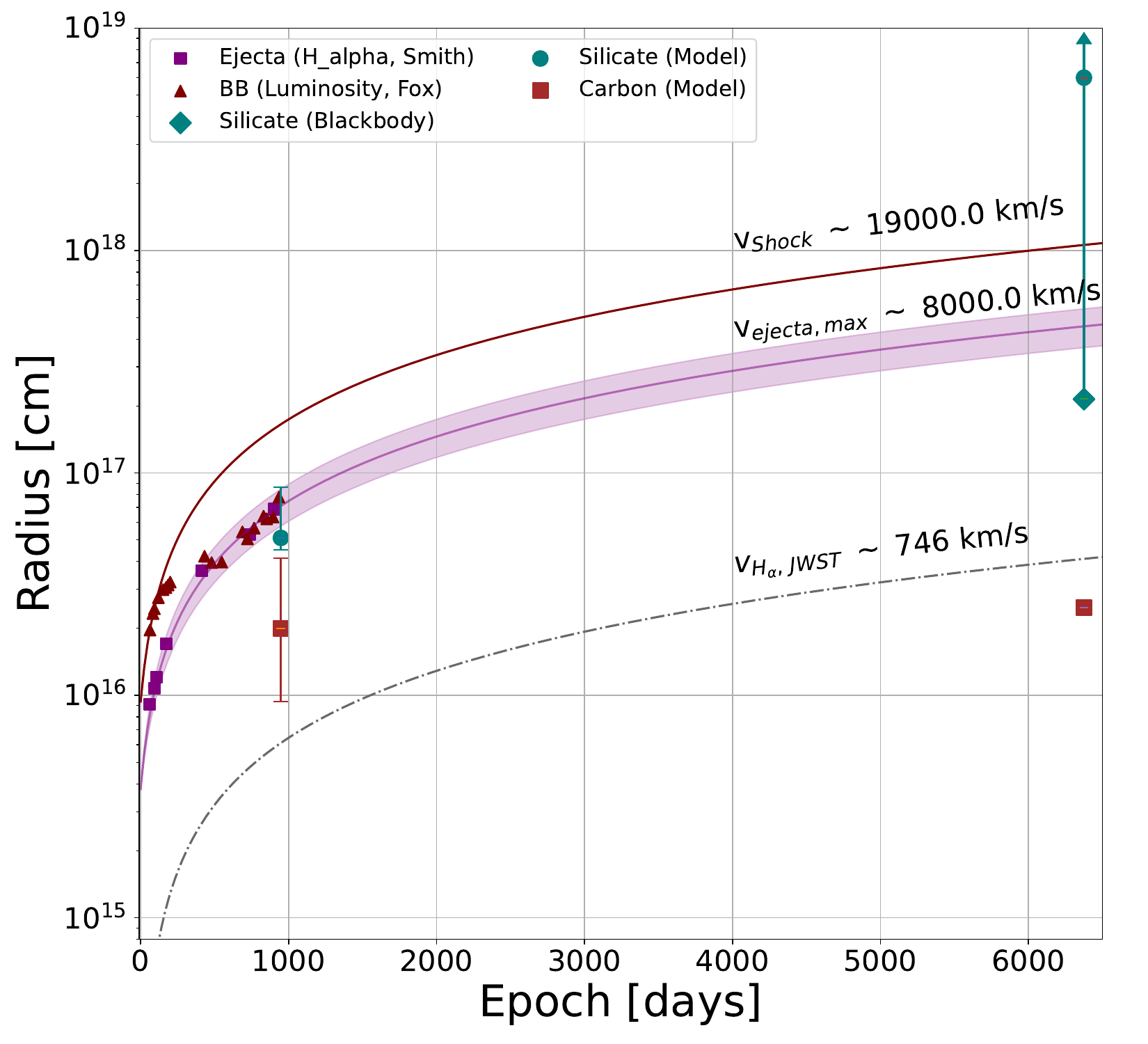}
    \caption{Plot of the modeled dust radii versus time for both the CD and WD components in the \sst~and \jwst~MIR spectra. The values for the radii are described in detail in Section \ref{sec:scenarios}. Overplotted are radii calculated for the forward shock radius derived from \citet{smith17}, as well as lines of constant velocity. These values are used to interpret the origin of the dust in SN 2005ip, described in Section \ref{subsec:formation}.}
    \label{fig:radii}
\end{figure}

\subsection{Dust Origin}\label{subsec:formation}
The dust geometry and evolution are important considerations for disentangling the dust origin. 
As previously noted, \citet{fox10} conclude the CD component is consistent with a pre-existing CSM that formed a spherical shell at or beyond the forward shock.
The latest \jwst~data, however, suggests that a pre-existing CSM does not likely make a strong contribution to the overall spectrum. 
This is because the \jwst~spectrum shows the CD component (and overall dust mass) is dominated by the optically thin Mg-silicate dust. 
In the \sst~spectrum, however, these Mg-silicate features are not visible. 
Instead, they are likely either non-existent or suppressed since the dust component was more compact and optically thick 15 years earlier. 
A scenario invoking a significant pre-existing CSM is unlikely since the CSM is static and cannot {\it both} increase in mass and decrease in optical depth over the 15-year gap between the two spectra.

We, therefore, consider other scenarios in which the dust forms in the ejecta and/or the CDS.
Figure \ref{fig:radii} plots the different radii for each component described at the beginning of Section \ref{sec:scenarios}. 
The figure also includes measurements of the radial evolution for lines of constant velocity consistent with both the ejecta and forward shock, as well as radii associated with measured shock velocities in \citet{smith19}. 
These points provide some useful constraints. 
First, the CD radii (silicate points) are too high to be consistent with the ejecta radii. 
The majority of the CD must not be located in the ejecta. 
Furthermore, the updated models place the CD radii within (not external to) the forward shock radius. 
For these reasons, we conclude CD in \ip\ is most likely forming inside the CDS. 
As the CDS expanded, the dust shell likely evolved from optically thick to optically thin while also continuing to form new dust. 
One additional consistency check is possible. 
As previously noted, Equation \ref{eqn:tau} shows that $\tau$~is inversely proportional to $R^2$. Figure \ref{fig:tau} shows that the CD component goes from optically thick, with an average $\tau\approx10$, to optically thin, suggesting an increase in radius by at least a factor of $3-4\times$. 
In fact, our lower limit on the CD radius increases by a factor of $\approx4$.

The WD graphite component is less significant in mass but may have a significant impact on our interpretation. 
For the WD component, the {\it minimum} dust radius is much smaller than the CD component and consistent with velocities seen in the slower-moving ejecta. 
It is important to note that this radius is only the lower limit. 
For any optically thin scenario, the upper limit has no bound. 
In the former scenario, dust would form in the expanding ejecta. 
In the latter scenario, the dust would be pre-existing in the pre-shocked CSM.
The precise location of the WD graphite component is not obvious, but we have some clues. 
First, the overall mass does not increase between the first and second epochs ($\sim$0.005 \msolar). 
This point may be interpreted as no new dust formation or that a nearly constant dust mass is continuously heated to an observable temperature. 
Second, both the WD composition and temperature are different from the CD, which suggests they are two distinct components. 
To ultimately disentangle the scenarios, it is useful to consider the heating mechanism.

\subsection{Dust Heating Mechanism}\label{subsec:heating}
While radioactive decay may be a viable heating mechanism at $<$1500 days, it is not sufficient to heat dust to the temperatures observed with the \jwst~data on day 6376. Even for the earlier \sst~data, previous analysis by \citet{fox10} rules out most possible heating scenarios except radiative heating from CSM interaction. In this scenario, the forward shock can generate a shock power of about $10^{40}$\,erg\,s$^{-1}$. A fraction of that power will come out as X-rays. The thermalized part of this power will emerge primarily in the UV, especially in Ly$\alpha$ 1215.67~\AA\ and Mg\,{\sc ii}\,2800~\AA, while a few percent of that thermalized flux emerges in the optical as a weak continuum source together with emission lines, in particular H$\alpha$ \citep{smith17a}. Although it wasn't clear from the \sst~data alone whether the observed dust was in the CDS or not, \citet{fox10} showed that the dust temperature and blackbody radius were consistent with the expected temperature from the CSM interaction scenario.

For this reason, the optical spectrum of the SN at very late times is critical, particularly for this analysis. 
The optical evolution of \ip, particularly the H$\alpha$~line, has been well documented \citep{smith09,stritzinger12,smith17,fox20}. Figure \ref{fig:optical} plots the most recent optical spectra
Compared to Figure~2 in \citet{fox20}, the spectrum has evolved very little. 
The integrated flux for the late-time optical spectra of \ip\ corresponds to only about $>10^5$~\lsolar.
Similar to the conclusions made in previous papers on \ip\ \citep[e.g.,][]{fox10,smith17,fox20}, the optical luminosity generated by the ongoing shock interaction is sufficient to heat the new dust. 
There may even be additional heating. \citet{dessart22} shows that the shock power introduced at the interface between ejecta and CSM emerges primarily in the UV, channeled primarily into Ly$\alpha$, for which we have no observational coverage, even in the archival {\it HST}/WFC3 UVIS data. 
These models suggest that the observed optical luminosities ($\sim 10^5$~\lsolar) could be compatible with a UV flux of $\sim 10^6$~\lsolar.

A useful representation of the likely geometry has been previously considered and illustrated in Figure~1 of \citet{smith17a}. 
The CD component lies within the CDS, while the WD is somewhere external to the CDS, either close to the forward or reverse shock. 
In the former scenario, the WD would be associated with pre-existing dust in the pre-shocked CSM. 
The hottest dust is closest to the forward shock and is continuously heated as the shock expands. 
Although the WD is carbon-rich, it does not preclude the presence of silicate-rich material in the CSM, given silicates have a much lower vaporization temperature ($\sim$1200 K) than the graphites ($\sim$1800 K). 
In the post-shock CDS, all dust compositions can then recondense. 
Alternatively, the WD component could be associated with newly formed dust in the ejecta and similarly heated by the reverse shock. 
In this scenario, the WD would have to be closer to the reverse shock, and the reverse shock would have to be nearly as strong as the forward shock.

Whether the radiation primarily comes from the forward or reverse shocks remains an open question.
The direction of the heating may affect the particular models that are used to fit the MIR spectra. 
We do not have X-ray or UV observations, but a large optical to X-ray/UV ratio may signify significant reprocessing of the shock radiation, which may suggest radiation from the reverse shock being reprocessed by the CDS. 
Reprocessing is possible for the forward shock, too, of course, although one may not expect an optical to X-ray/UV ratio as high. 
Alternatively, the H$\alpha$~profile in Figure \ref{fig:optical} does not have any broad component associated with the ejecta, which may suggest we are predominantly seeing only the forward shock. 

Other possible scenarios exist for the heating of the dust. 
For example, the majority of the flux may come out in X-rays at late times. 
The reverse shock may have traveled sufficiently far back into the ejecta to heat them directly (i.e., not radiatively). 
The optical depth is quite high, thereby absorbing 99\% of the optical flux, or there may be a pulsar at the center. 
Each of these scenarios requires additional observations and more complex models.

\begin{figure}
    \centering
    \includegraphics[width=0.49\textwidth]{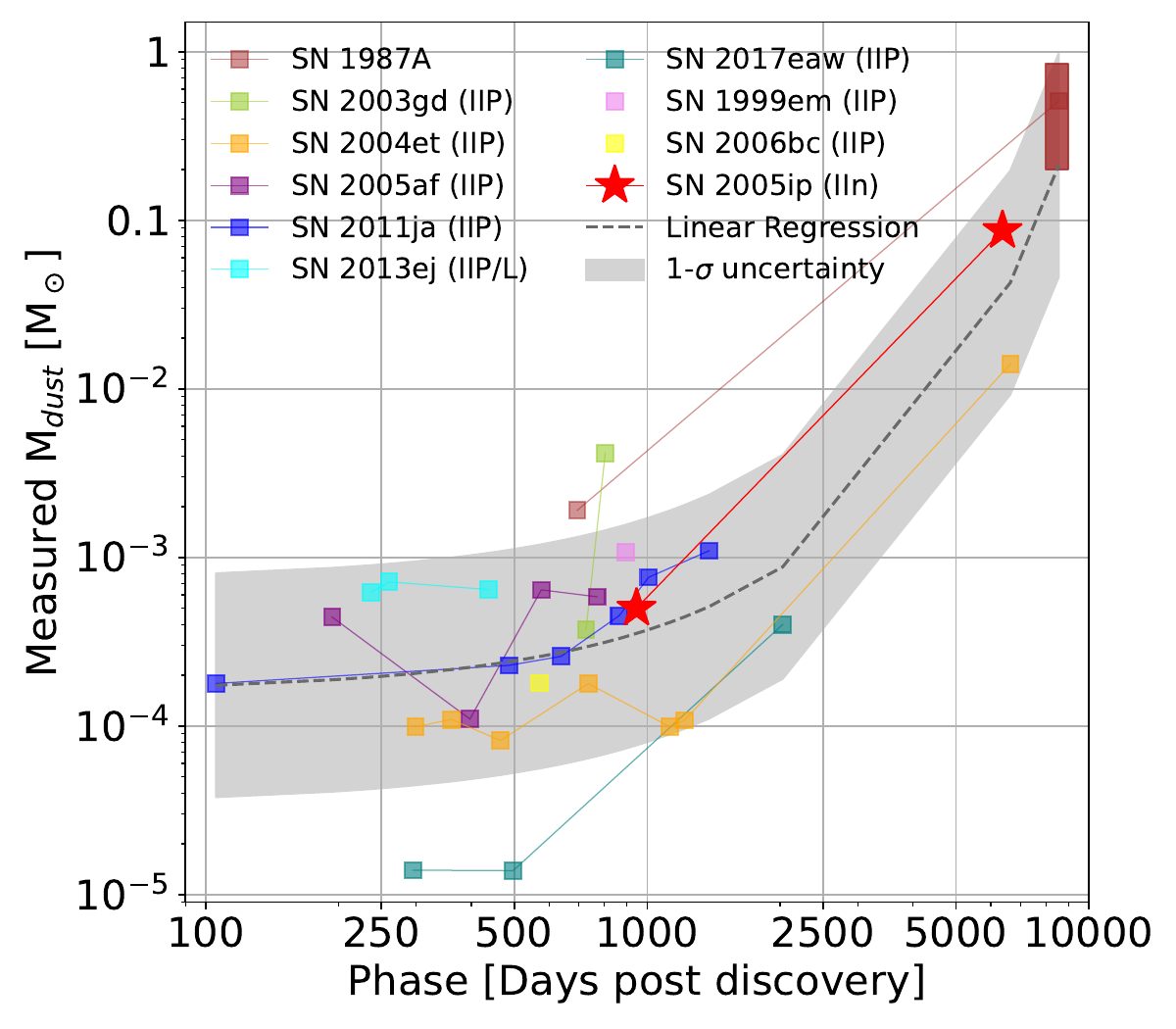}
    \caption{The dust mass in \ip\ as a function of the epoch of MIR observations compared with other historic dusty SNe. Overall, compared to other historical SNe, the inferred dust mass in \ip\ is one of the highest to date.}
    \label{fig:all_SNe_dmass}
\end{figure}

\section{Conclusions} \label{sec:conclusions}
This paper presents \jwst/MIRI MRS observations of \ip, which were conducted 6376 days post-discovery. 
\ip\ is a well-studied Type IIn SN that has continued to show signatures of dust and shock interaction for nearly 20 years post-explosion. 

The primary takeaway is that two epochs of MIR spectra spaced nearly fifteen years apart show clear evolution, revealing an increasingly large cold dust component totaling nearly $\sim$0.1 \msolar (and a much less massive warm dust component). 
Specifically, a Mg-silicate feature is present in the \jwst spectrum that is not detected in the earlier \sst spectrum. 
This is strong evidence that the cold dust component was more compact and optically thick on day $\sim$ 1000 or simply too small to be detected. 
Either way, the majority of this cold dust component was not pre-existing.

A detailed analysis of the cold dust component places the dust roughly in the radius of the forward shock, leading us to conclude that the new dust forms in the CDS behind the forward shock. 
This is consistent with previous suggestions based on line profile shapes \citep{smith09}. 
A contemporaneous optical spectrum exhibits a powerful H$\alpha$~line ($\sim10^5$~\lsolar), indicating sufficient shock interaction to radiatively heat the dust to the observed temperature and luminosity.

Overall, compared to other historical SNe, the inferred dust mass in \ip\ is one of the highest to date (Figure \ref{fig:all_SNe_dmass}).
The formation of significant dust in the CDS presents an alternative to the SN ejecta, which has not produced the expected yields in SNe~IIP \citep[e.g.,][]{kozasa09a,meikle11}. 
Although SNe~IIn make up less than 10\% of the CCSN population in the low-redshift Universe \citep{li11,smith11}, they may be more common at high redshift. A top-heavy IMF with more high-mass stars, similar to LBV progenitors, and a higher fraction of binary stars that undergo stripping \citep{doughty21} could lead to more Type IIn-like environments with dense CSM, resulting in significant shock interaction and a CDS.
Even so, \ip\ is just one object and an extreme example of the SN~IIn subclass. 
A larger sample of SNe~IIn observed in the MIR with long baselines is ultimately needed to determine the contribution of SNe~IIn to the overall galactic dust budget.

\section*{Acknowledgements}
M.S. is supported by an STScI Postdoctoral Fellowship.
M.S. acknowledges support by NASA/JWST grants GO-04436, GO-04217, GO-01860, and GO-02666. T.T acknowledges support by NSF grant AST-2205314.
This project was supported in part by the Transients Science @ Space Telescope group.
This work was supported by a NASA Keck PI Data Award, administered by the NASA Exoplanet Science Institute. The Observatory was made possible by the generous financial support of the W. M. Keck Foundation. The authors wish to recognize and acknowledge the very significant cultural role and reverence that the summit of Maunakea has always had within the indigenous Hawaiian community. We are most fortunate to have the opportunity to conduct observations from this mountain.
This work is based on observations made with the NASA/ESA/CSA {\it James Webb Space Telescope}. Data were obtained from the Mikulski Archive for Space Telescopes at the Space Telescope Science Institute, which is operated by the Association of Universities for Research in Astronomy, Inc., under NASA contract NAS 5-03127 for {\it JWST}. These observations are associated with program \#1860.
Some of the data presented herein were obtained at the W. M. Keck Observatory, which is operated as a scientific partnership among the California Institute of Technology, the University of California, and NASA; the observatory was made possible by the generous financial support of the W. M. Keck Foundation. 
A.V.F.'s supernova group at U.C. Berkeley is grateful for financial assistance from the Christopher R. Redlich Fund,
Gary and Cynthia Bengier, Clark and Sharon Winslow, Alan Eustace (W.Z. is a Bengier-Winslow-Eustace Specialist in Astronomy), William Draper, Timothy and Melissa Draper, Briggs and Kathleen Wood, Sanford Robertson (T.G.B. is a Draper-Wood-Robertson Specialist in Astronomy), and many other donors.
This project has been supported by the
NKFIH OTKA FK-134432 grant of the National Research, Development, and
Innovation (NRDI) Office of Hungary. S.Z. is supported by the ÚNKP-23-4-
SZTE-574 New National Excellence Program of the Ministry for Culture and
Innovation from the source of the NRDI Fund, Hungary.

\section*{Data Availability}
The \jwst\ raw data associated with this program can be found at
\dataset[DOI:10.17909/q96n-2296]{\doi{10.17909/q96n-2296}}.

\facilities{\jwst\ (MIRI/MRS), \hst\ (STIS), Swift (XRT and UVOT), AAVSO, CTIO:1.3m,
CTIO:1.5m,CXO}

\software{astropy \citep{theastropycollaboration13}}

\bibliographystyle{aasjournal}
\bibliography{05ip_references}
\end{document}